\documentclass[amsmath,amssymb,12pt,floatfix] {revtex4}
\usepackage[dvips]{graphicx}
\usepackage{bm}

\newcommand {\ber}{\begin{eqnarray}}
\newcommand {\eer}{\end{eqnarray}}
\newcommand{\be}{\begin{equation}}
\newcommand{\ee}{\end{equation}}
\newcommand{\bel}[1]{\begin{equation}\label{#1}}
\newcommand{\bea}{\begin{eqnarray}}
\newcommand{\eea}{\end{eqnarray}}
\newcommand{\ba}{\begin{array}}
\newcommand{\ea}{\end{array}}

\newcommand{\bfx}{\mathbf{x}}
\newcommand{\bfy}{\mathbf{y}}

\def\Z{{\mathbb Z}}

\begin{document}
\title{Determinant representation for some transition probabilities in the TASEP with second class particles}
\author{Sakuntala Chatterjee(1,2) and Gunter M. Sch\"{u}tz(1) }
\affiliation{ (1) Institut f\"{u}r Festk\"{o}rperforschung, 
Forschungzentrum J\"{u}lich, D-52425 J\"{u}lich, Germany. \\
(2) Physics Department, Technion, Haifa-32000, Israel.}
\begin{abstract}
Abstract: We study the transition probabilities for the totally asymmetric 
simple exclusion process (TASEP) on the infinite integer lattice with a
 finite, but arbitrary number of first and second class particles. 
Using the Bethe ansatz we present an explicit expression of these
quantities in terms of the Bethe wave function. In a next step it is proved 
rigorously  that this expression can be written in a compact determinantal
 form for the case where the order of the 
first and second class particles does not change in time. An independent
 geometrical approach provides insight into these results and enables us 
to generalize the determinantal solution to the multi-class TASEP.
\end{abstract}
\maketitle
\section{Introduction}
\label{sec:intro}

In this paper we derive finite-time transition probabilities for the 
 one-dimensional totally
asymmetric simple exclusion process (TASEP) with several classes (species)
 of particles
\cite{Ligg99,Schu01}. In the usual single-species TASEP in one dimension
 particles jump on the integer lattice $\Z$ 
independently after an exponentially distributed random time with parameter 1 
to their right nearest neighbor site, provided that the
target site is empty. Otherwise the jump attempt is rejected.
This exclusion principle guarantees that each site is always occupied 
by at most one
particle. An instantaneous configuration of this system with $N$ particles can therefore
be represented by an ordered set of integer coordinates $\{\bfx\}=\{x_1,\dots,x_N\}$
where $x_{i+1}>x_i$. 
An intuitive form to represent this lattice gas dynamics consists in writing allowed
local transitions as
\bea
A0 & \to & 0A
\eea
where $A$ represents a particle and $0$ represents a vacant site.
Mathematically, the dynamics of the TASEP can be defined through a 
master equation 
for the probability $P(\bfx,t)$ to find a configuration $\bfx$ at time $t$. 
The master equation reads
\bel{1-1}
\frac{d}{dt} P(\mathbf{x},t) = \sum_{\bfx'\neq \bfx} 
 \left[w_{\bfx,\bfx'}  P(\mathbf{x}',t) - w_{\bfx',\bfx}  P(\mathbf{x},t) \right],
\ee
where $w_{\bfx',\bfx}$ is the transition rate (0 or 1 for the TASEP) to
go from a configuration ${\bf x}$ to a configuration ${\bf x}^{'}$.
By integrating (\ref{1-1}) one obtains the solution of the master equation for any given
initial configuration $\bfy$, i.e., the conditional probability to find a particle configuration $\bfx$ at time $t$,
given that the process started from configuration $\bfy$. 

This process  is a discretised version of the noisy Burgers 
equation~\cite{Spoh91,burgers} 
or, equivalently, the noisy Kardar-Parisi-Zhang equation \cite{Kard86,Sasa10}.
On coarse-grained scale the Burgers equation exhibits shock solutions and it is
of great interest to understand the microscopic structure of the shocks on the
scale of the lattice. To this end, one introduces so-called second-class particles \cite{Ferr91}.
We denote the first class particles by the symbol
$A$ and the second class particles by the symbol $B$. 
 According to the exclusion principle a lattice site can either be occupied by an  $A$
particle, or a $B$ particle, or it can remain 
vacant ({\sl i.e.} occupied by a hole, denoted as $0$). The exchange rules
are
\begin{eqnarray}
\nonumber
A0 &\rightarrow& 0A \\
B0 &\rightarrow& 0B \\
\label{eq:rate}
\nonumber
AB &\rightarrow& BA.
\end{eqnarray}
All processes occur with rate unity.  Observe that from the perspective 
of the first-class particles the second-class particles are not
 distinguishable from holes. The stationary distribution of this two-component
TASEP can be constructed using pair of independent product
measures~\cite{angel}.
It is well known that a single second-class particle in a system of 
first-class particles travels
with the characteristic speed of the Burgers equation. Hence it can be used to define
a microscopic position of the shock and to study its microscopic structure 
\cite{Ferr91,Ferr94,Derr97a,Beli02,Kreb03,speer}.  It was also
shown that a finite number of second class particles, in a uniform 
stationary background
of first class particles, form a weak bound state~\cite{Derr93c}.   
 The TASEP with
second-class particles and more general two-component particle systems are
 also 
interesting from the point of view of non-equilibrium phase transitions
\cite{godreche,gms2003,blythe}.

More generally, one may consider a
hierarchical multi-class particle system where for a particle of 
class $k$ all particles
of higher class behave like holes. This has interesting applications 
e.g. in the microscopic 
study of the so-called step initial  condition where in an infinite
 system all sites $x\leq 0$ are 
occupied while the positive half lattice $x>0$ is vacant. This initial
 condition has been studied in
great detail for the single-species TASEP, see e.g. 
the work by Johansson \cite{Joha00} for a seminal study of current 
fluctuations.
Recently, interesting passing probabilities for the multi-class 
exclusion process have been
obtained~\cite{patricia}. The stationary distribution of the multi-class 
TASEP has
been constructed in \cite{jbmartin} by extending the approach of \cite{angel}.
A matrix product formalism for the stationary measure of a multi-class TASEP was
discussed in \cite{kirone}.

A determinantal approach to solve for the dynamics of the single-species
 TASEP  on the 
infinite lattice has been introduced in~\cite{Schu97}. It is based
on the explicit solution of the master equation for the conditional
probability of finding the system in a certain configuration at time $t$ for a
given initial configuration, using the technique of the Bethe Ansatz, originally employed 
for studying the relaxation spectrum of a periodic chain \cite{Gwa92}. 
Moreover, it was shown that the solution of master equation for $N$
particles on an infinite lattice can be written as the determinant of  
an $N \times N$  matrix. This determinantal form has allowed for a 
rederivation and extension of
the work of Johansson on the step initial condition 
\cite{Naga04,Rako05,Prie08} and has inspired a substantial
body of further work on other dynamical properties of the exclusion 
process, see e.g. 
\cite{Bran04,Sasa05,parallel,Sasa07,Boro07,Boro08,Tracy1,Tracy2}.
In the present work we obtain an exact solution of the master equation 
for a finite number of 
first class and second class particles on an infinite lattice with
 arbitrary initial condition. We use 
the {\it nested Bethe ansatz} \cite{schultz,yang,suth}, thus generalizing 
the approach of \cite{Schu97} for the single-species 
case. In a recent study the technique of nested Bethe ansatz was used to
investigate the spectral structure of the transition matrix for a
multi-species exclusion process \cite{arita}. In particular, the relaxation
time-scale of the system was studied and it was shown that the dynamical
exponent remains same as in the single-species case. In this paper,
we use nested Bethe
ansatz to solve the master equation of the two-species case. 
 We obtain a compact determinantal representation of the
solution for the case where first and second-class particles do not 
change their ordering. We also discuss the extension of our results to 
multi-species exclusion processes.    

The paper is organized as follows.
In Sec. \ref{sec:bethe} we show how the master equation 
can be solved exactly for the two-species case using the nested Bethe
 ansatz. In section \ref{sec:theorem} we 
give a formal proof of the determinantal representation for transition
 probabilities without interchange of particles.
 In section \ref{sec:dia} we
present an alternative diagrammatic approach  and
discuss how using this approach the determinantal representation can be
understood. We conclude this paper by discussing the possible generalization
 to multi-species systems in section \ref{sec:last}.

\section{Exact solution of the master equation using Bethe Ansatz}
\label{sec:bethe}

In this section we present an exact solution of the master equation for the
TASEP with first and second class particles. To illustrate the approach 
we first consider the case of total particle number $N=2$. Each of these
particles can be either of type $A$ or of type $B$. Therefore there are $4$
different particle sequences: $\{AA, AB, BA, BB \}$. 

Let $x_1, x_2$ with $x_2 \geq
x_1+1$ be the positions of the two particles at time $t$ and 
$y_1,y_2$ be their initial positions with $y_2 \geq y_1+1$. The quantity
$P^{Q'|Q}\left ( x_1,x_2;t|y_1,y_2;0 \right ) $ denotes the probability to
find the system in sequence $Q' \in  \{AA, AB, BA, BB \}$ with occupancies at
$x_1, x_2$ at time $t$, given that at $t=0$ the sequence was $Q \in  \{AA, AB,
BA, BB \}$ with occupied sites at $y_1,y_2$. Note that if the initial sequence
is $AA, BA$ or $BB$, then no interchange between $A$ and $B$ particles is
possible as the system evolves in time. For these cases the sequence remains
the same for all times and the particles execute usual TASEP dynamics with transition
probabilities obtained in~\cite{Schu97}. However, if the initial sequence
is $AB$, then as a result of interchange between the $A$ and $B$ particles
the sequence may change and the possible sequences at time $t$ are $AB, BA$.  

In the absence of nearest neighbor pairs, {\sl i.e.} for $x_2 > x_1+1$, the
master equation becomes
\begin{eqnarray}
\nonumber
\frac{d }{dt}P^{Q'|Q}(x_1,x_2;t|y_1,y_2;0) =  &P^{Q'|Q}(x_1-1,x_2;t|y_1,y_2;0)
+  P^{Q'|Q}(x_1,x_2-1;t|y_1,y_2;0)
\\
  & -2P^{Q'|Q}(x_1,x_2;t|y_1,y_2;0). 
\label{eq:master}
\end{eqnarray}
For $x_2 = x_1+1$, the form of the equation is different depending on $Q$ and
$Q'$. For $Q \in \{AA,BA,BB\}$ we have $Q'=Q$ and the equation takes the form
\begin{equation}
\frac{d }{dt}P^{Q|Q}(x_1,x_1+1;t|y_1,y_2;0)=P^{Q|Q}(x_1-1,x_1+1;t|y_1,y_2;0)
- P^{Q|Q}(x_1,x_1+1;t|y_1,y_2;0).
\label{eq:bdry1}
\end{equation}
For $Q=AB$ and $Q'=AB$ we have 
\begin{equation}
\frac{d }{dt}P^{AB|AB}(x_1,x_1+1;t|y_1,y_2;0) = P^{AB|AB}(x_1-1,x_1+1;t|y_1,y_2;0)
-2P^{AB|AB}(x_1,x_1+1;t|y_1,y_2;0)
\label{eq:bdry2}
 \end{equation}
while for $Q=AB$ and $Q'=BA$
\begin{eqnarray}
\nonumber
\frac{d }{dt}P^{BA|AB}(x_1,x_1+1;t|y_1,y_2;0) = & P^{BA|AB}(x_1-1,x_1+1;t|y_1,y_2;0)
+  P^{AB|AB}(x_1,x_1+1;t|y_1,y_2;0) \\
&-P^{BA|AB}(x_1,x_1+1;t|y_1,y_2;0).
\label{eq:bdry3}
 \end{eqnarray}

To solve the above set of equations for a given initial sequence $Q$ and
initial positions $\{y_1,y_2\}$ we define 
\begin{equation}
\left | P(t) \right \rangle = \left ( 
\begin{array}{c}
P^{AA|Q} (x_1,x_2;t|y_1,y_2;0) \\
P^{AB|Q} (x_1,x_2;t|y_1,y_2;0) \\
P^{BA|Q} (x_1,x_2;t|y_1,y_2;0) \\
P^{BB|Q} (x_1,x_2;t|y_1,y_2;0) 
\end{array}
\right ).
\end{equation}
Notice that in order to keep the
notation simple we do not mark the dependence of this vector on the
initial sequence $Q$.
Now we use Bethe Ansatz to write $\left | P(t) \right \rangle $ in the
following form
\begin{equation}
\left | P(t) \right \rangle = \frac{1}{\left ( 2 \pi \right )^2} \int_0
^{2\pi } \int_0 ^{2\pi } dp_1 dp_2 e^{-\epsilon \left ( p_1, p_2 \right ) t}
\left ( e^{i \left ( p_1x_1 + p_2 x_2 \right )} + \Sigma \left
(p_1, p_2 \right )  e^{i \left ( p_2x_1 + p_1 x_2 \right )} \right ) \left |
P(0) \right \rangle
\end{equation}
where 
\begin{equation}
\left | P(0) \right \rangle = \left ( 
\begin{array}{c}
\delta_{AA,Q} \\
\delta_{AB,Q} \\
\delta_{BA,Q} \\
\delta_{BB,Q} 
\end{array}
\right )
e^{-i\left ( p_1y_1 + p_2y_2 \right )}.
\end{equation}
The elements of the $4 \times 4$ matrix $\Sigma(p_1,p_2)$ and the quantity $\epsilon
\left ( p_1, p_2 \right )$ are yet undetermined functions of $p_1$ and $p_2$. 
Substituting this ansatz into (\ref{eq:master}) we find
\begin{equation}
\epsilon (p_1, p_2) = 2 - e^{-ip_1} -e^{-ip_2}
\end{equation}
and to satisfy the boundary conditions for the 
nearest neighbor pairs resulting from Eqs (\ref{eq:bdry1}), (\ref{eq:bdry2}),
(\ref{eq:bdry3}), one needs 
\begin{equation}
\label{Sigmamatrix}
\Sigma (p_1,p_2) = \left ( 
\begin{array}{cccc}
-\frac{1-e^{ip_2}}{1-e^{ip_1}} & 0 & 0 & 0 \\
0 &  -1 & 0 & 0 \\
0 & -\frac{e^{ip_1}-e^{ip_2}}{1-e^{ip_1} } & -\frac{1-e^{ip_2}}{1-e^{ip_1}} &
0 \\
0 & 0 & 0 &  -\frac{1-e^{ip_2}}{1-e^{ip_1}} 
\end{array}
\right ).
\end{equation} 
The complete solution can then be written in terms of the functions 
\be
F_n(x;t) = \frac{1}{2\pi} \int_0 ^{2 \pi} dp e^{-t \left ( 1-e^{-ip} \right )}
\frac{e^{ipx}}{\left ( 1-e^{i \left (p+i0 \right )} \right ) ^n}. 
\label{eq:fnx}
\ee   
introduced in~\cite{Schu97}.
From now on we will drop $t$ from the argument of the above function
and write it simply as $F_n(x)$, for
notational simplicity. 

For the present case of two particles the values
 of $n$ that occur are $0, \pm 1$. For $Q = AA, BA, AB$ we have 
\be 
P^{Q|Q} \left (x_1,x_2;t|y_1,y_2,0 \right ) = F_0 (x_1 -y_1)F_0 (x_2-y_2) -
F_1(x_2-y_1)F_{-1}(x_1 -y_2)
\ee
and all other elements of $\left | P(t) \right \rangle$ with $Q' \neq Q$ are
zero. For $Q=AB$ we have
\be
\left | P(t) \right \rangle = \left ( 
\begin{array}{c}
0 \\
P^{AB|AB} \left (x_1,x_2;t|y_1,y_2,0 \right )\\
P^{BA|AB} \left (x_1,x_2;t|y_1,y_2,0 \right ) \\
0
\end{array}
\right ) = 
\left ( 
\begin{array}{c}
0 \\
F_0 (x_1 -y_1)F_0 (x_2-y_2) - F_0(x_2-y_1)F_{0}(x_1 -y_2) \\
F_0 (x_2 -y_1)F_0 (x_1-y_2) - F_1(x_2-y_1)F_{-1}(x_1 -y_2) \\
0
\end{array}
\right )
\label{eq:vr2}
\ee

The important point to note for $N \geq 3$ is that there are no new constraints from the
boundary condition when more than two particles are on adjacent sites. This is expected from
the integrability of the model \cite{Alca93,Popk02}, but we have also
verified this explicitly in our calculation (see appendix). For $N=3$ there
are $8$ possible sequences of $A$ and $B$, ordered in a vector 
$\left | P(t) \right \rangle$ according to the
following formal tensor product:
\be
\left ( \begin{array}{c} 
A\\
B
\end{array}\right ) ^{\otimes 3} =
\left ( \begin{array}{c}
AAA\\
AAB\\
ABA \\
ABB \\
BAA\\
BAB\\
BBA\\
BBB
\end{array}\right ).
\ee
The Bethe Ansatz solution has the form 
\ber
\nonumber
|P(t) \rangle =\frac{1}{(2 \pi)^3} \int_0 ^{2 \pi} 
  dp_1 dp_2 dp_3 
e^{  -\epsilon (p_1, p_2) t }  \{
 e^{i(p_1x_1+p_2x_2+p_3x_3)} 
\\
\nonumber
+ \Sigma_{12}^{3}(p_1,p_2) e^{i(p_2x_1+p_1x_2+p_3x_3)} 
+ \Sigma_{23}^{3}(p_2,p_3)  e^{i(p_1x_1+p_3x_2+p_2x_3)}
\\
\nonumber
+\Sigma_{23}^{3}(p_1,p_3) \Sigma_{12}^{3}(p_1,p_2) 
e^{i(p_2x_1+p_3x_2+p_1x_3)} 
+\Sigma_{12}^{3}(p_1,p_3) \Sigma_{23}^{3}(p_2,p_3)
 e^{i(p_3x_1+p_1x_2+p_2x_3)}
\\
+\Sigma_{23}^{3}(p_1,p_2)\Sigma_{12}^{3}(p_1,p_3)
 \Sigma_{23}^{3}(p_2,p_3)  e^{i(p_3x_1+p_2x_2+p_1x_3)} \} |P(0) \rangle 
\label{eq:bethe3}
\eer
where $\Sigma_{12}^{3}(p_j,p_k)=\Sigma (p_j,p_k) \otimes {\mathbb 
I}_2 $ and $\Sigma_{23}^{3}(p_j,p_k) = {\mathbb I}_2 \otimes
\Sigma (p_j,p_k) $. Here ${\mathbb I}_2$ is the $2 \times 2 $ unit
matrix. Here the upper index 3 refers to the number of particles $N=3$. 
One crucial requirement for the integrability of the model is that these 
$\Sigma$-matrices must satisfy Yang-Baxter criterion (see appendix for
details).

Using the definition in (\ref{eq:fnx}) the above solution can be
written in terms of $F_n (x_i -y_j)$ with $n=0,\pm 1, \pm 2$. For example, for the
initial sequence $Q=AAB$ we find (with $\bfx=( x_1,x_2,x_3),\,\bfy=(y_1,y_2,y_3)$)
\be
P^{AAB|AAB} \left (\bfx ;t | \bfy ;0 \right ) = \left |
\begin{array}{ccc}
F_0 (x_1 -y_1) & F_{-1} (x_1 -y_2) & F_{-1} (x_1 -y_3) \\
F_{1} (x_2 -y_1) & F_0 (x_2 -y_2) & F_0 (x_2 -y_3) \\
F_{1} (x_3 -y_1) & F_0 (x_3 -y_2) & F_0 (x_3 -y_3) 
\end{array}
\right |
\ee
\ber
\nonumber
P^{ABA|AAB} \left ( \bfx ;t|\bfy ;0 \right ) = F_0 (x_1 -y_1)
\left \{ F_0(x_2-y_3)F_0(x_3-y_2) - F_1(x_3-y_2)F_{-1}(x_2-y_3) \right \} 
\\
\nonumber
+F_{-1} (x_1 -y_2)
\left \{ F_2(x_3-y_1)F_{-1}(x_2-y_3) - F_0(x_2-y_3)F_{1}(x_3-y_1) \right \}
\\
\nonumber
+F_{0} (x_2 -y_1)
\left \{ F_1(x_3-y_2)F_{-1}(x_1-y_3) - F_0(x_1-y_3)F_{0}(x_3-y_2) \right \} 
\\
+F_{-1} (x_2 -y_2)
\left \{ F_1(x_3-y_1)F_{0}(x_1-y_3) - F_2(x_3-y_1)F_{-1}(x_1-y_3) \right \} 
\eer 
\ber
\nonumber
P^{BAA|AAB} \left ( \bfx ;t|\bfy ;0 \right )  =  F_1 (x_3 -y_2)
\left \{ F_1(x_2-y_1)F_{-2}(x_1-y_3) - F_0(x_2-y_1)F_{-1}(x_1-y_3) \right \} 
\\
\nonumber
 +F_{0} (x_3 -y_2)
\left \{ F_0(x_1-y_3)F_{0}(x_2-y_1) - F_1(x_2-y_1)F_{-1}(x_1-y_3) \right \} 
\\
\nonumber
+F_{2} (x_3 -y_1)
\left \{ F_0(x_2-y_2)F_{-2}(x_1-y_3) - F_{-1}(x_2-y_2)F_{-1}(x_1-y_3) \right
\}  
\\
+F_{1} (x_3 -y_1)
\left \{ F_{-1}(x_2-y_2)F_{0}(x_1-y_3) - F_0(x_2-y_2)F_{-1}(x_1-y_3) \right
\}
\eer
and all other components of  $\left | P(t) \right \rangle$ are zero. Thus, starting
from any given initial sequence $Q$, all components of  $\left | P(t) \right
\rangle$ can be determined. 

Similarly, for $N=4$, the  possible 16 sequences can be represented 
in the tensorial ordering
$ \left ( \begin{array}{c} 
A\\
B
\end{array}\right ) ^{\otimes 4} $. The Bethe Ansatz solution can be written
down in terms of the matrices
\ber
\Sigma_{12}^{4} \left ( p_i, p_j\right ) &= & \Sigma \left ( p_i,
p_j\right ) \otimes {\mathbb I}_2 \otimes {\mathbb I}_2 \\
\Sigma_{23}^{4} \left ( p_i, p_j\right ) &= & {\mathbb I}_2 \otimes
\Sigma \left ( p_i, p_j\right ) \otimes {\mathbb I}_2  \\
\Sigma_{34}^{4} \left ( p_i, p_j\right ) &= & {\mathbb I}_2 \otimes  
{\mathbb I}_2 \otimes \Sigma \left ( p_i, p_j\right ) 
\eer
and in this way one can construct $\left | P(t) \right \rangle$ for any 
value of $N$. The total of  number of sequences is  $2^N$ and
these sequences are ordered as in $$\left ( 
\begin{array}{c} 
A\\
B
\end{array}\right ) ^{\otimes N} $$  The $2^N$ dimensional $\Sigma$-matrices
are constructed from the tensor product of $\Sigma (p_i, p_j)$ and 
${\mathbb I}_2$ matrices: 
\ber
\Sigma_{12}^N (p_i,p_j) & = &\Sigma (p_i, p_j) \otimes {\mathbb I}_2 \otimes
{\mathbb I}_2 \otimes .....\otimes  {\mathbb I}_2 \\
\Sigma_{23}^N (p_i,p_j)  & = & {\mathbb I}_2 \otimes \Sigma (p_i, p_j) \otimes
{\mathbb I}_2 \otimes .....\otimes  {\mathbb I}_2 
\eer
etc. and in general 
\ber
\Sigma_{k,k+1}^N (p_i,p_j)  & = & \left ( {\mathbb I}_2 \right ) ^{\otimes
k-1}  \otimes \Sigma (p_i, p_j) \otimes \left ( {\mathbb I}_2 \right )
 ^{\otimes N-1-k}
\label{eq:nsigma}
\eer
The conditional probability $P^{Q'|Q} \left (
x_1,x_2,...,x_N;t|y_1,y_2,...,y_N;0 \right ) $ has the Bethe ansatz form
\be
\left | P(t) \right \rangle = \frac{1}{(2\pi)^N} \int_0^{2\pi} dp_1
dp_2...dp_N \sum_{\{Q\}} {\cal P} (Q_1Q_2...Q_N) e^{i\left
(p_{Q_1}x_1+p_{Q_2}x_2+...+p_{Q_N}x_N \right )} \left | P(0) \right \rangle.
\label{eq:nbethe}
\ee 
In this Bethe representation $\{Q_1Q_2...Q_N\}$ is one particular
 permutation of $\{12...N\}$  and
${\cal P} (Q_1Q_2...Q_N)$ is constructed from the product of $\Sigma_{i,i+1}^N
\left ( p_k, p_j\right )$ matrices given by Eq. (\ref{eq:nsigma}). 
This product corresponds to the series of 
elementary permutations performed on the momenta $\{p_1,p_2,...,p_N\}$ to
arrange them in the order $\{p_{Q_1},p_{Q_2},...,p_{Q_N}\}$. The summation 
is over all possible $N!$ permutations. One can 
 show  by adapting the techniques of \cite{Popk02} that Eq.
(\ref{eq:nbethe}) solves the $N$-particle master equation for the TASEP 
with second  class particles.

Notice that the matrix $\Sigma$  in Eq. (\ref{Sigmamatrix}) may be decomposed as 
$\Sigma=D+A$ where $D$ is the diagonal part and $A$ the off-diagonal part which 
arises from the interchange $AB \to BA$ of the two species of particles. 
Correspondingly the product ${\cal P} (Q_1Q_2...Q_N)$ can be expanded in terms
containing an increasing number of $A$ matrices which represent the number
of interchanges in the final configuration. The purely diagonal part in this expansion
corresponds to configurations in which the original order of first and second-class
particles has not changed. We investigate these realizations of the stochastic
dynamics in more detail in the following section.

\section{Theorem for no interchange}
\label{sec:theorem}

In the previous section we have shown that using Bethe Ansatz it is possible
to solve the master equation for any number of particles with any sequence.
The Bethe Ansatz form of the solution for $N$ particles has, in general, $N!$
number of terms.  In the single-species TASEP case, it was shown
in~\cite{Schu97} that the solution of the master equation can be written as a
determinant of a matrix $F$ whose elements are $F_{ij}=F_{i-j}(x_i-y_j)$ with
$F_{i-j}(x_i-y_j)$ defined according to (\ref{eq:fnx}).  In this section we
show that there is compact determinant representation of the transition probability
for the case where
no interchange between $A$ and $B$ particles takes place, i.e., where the sequence at
time $t$ is the same as the initial sequence at time $t=0$.  We state the main result:\\

\noindent {\bf Theorem for no interchange:} Consider a system of $N$ particles, each of
species $A$ or $B$, starting with an initial sequence  $Q$  at time $t=0$ taken from the tensorial
ordering $$\left ( 
\begin{array}{c} 
A\\
B
\end{array}\right )^{\otimes N}. $$ The probability 
$P^{Q|Q} \left ( x_1,x_2,...,x_N;t|y_1,y_2,...,y_N;0 \right ) $ where the
initial and final sequence remains the same can be written as the determinant of the
$N \times N$ matrix $G$, where 
\be
G_{ij} = F_{n(i,j)} \left ( x_i - y_j \right ) 
\ee
with matrix elements $F_n$ defined in (\ref{eq:fnx}) and 
 $n(i,j) = sgn(i-j) \left ( |i-j| -n_{AB} \right ) $ with $n_{AB}$ being
the number of $AB$ pairs occurring between the $i$-th and $j$-th
particle of the sequence $Q$. \\

Remark: To illustrate this result, to be proved rigorously below, consider e.g. the sequence
$AABABBAB...$. Here $n(1,3)=-1$, $n(1,7)=-4$, $n(8,4)=2$, $n(3,3)=0$, and so
on. Note that if a sequence $Q$ has no $AB$ pair, $n_{AB}=0\, \forall\, i,j$ and the
matrix $G$ becomes identical to the matrix found in \cite{Schu97}. This is expected 
since the system behaves in this case like
a single-species TASEP.

{\bf Proof:} We have to show first that $\det G$ satisfies the master equation. In the
absence of any nearest neighbor pair, the master equation is 
\ber
\nonumber
\frac{d}{dt} P^{Q|Q} \left (\bfx;t|\bfy;0 \right ) = 
\sum_{i=1}^N P^{Q|Q} \left (x_1,x_2,...,x_{i-1},
x_{i}-1,x_{i+1},...,x_N;t|\bfy;0 \right ) 
\\
- N P^{Q|Q} \left
(\bfx;t|\bf y;0 \right ). 
\label{eq:qqmaster}
\eer 
Since all the elements of the matrix $G$ contain the factor $e^{-t}$, one can
write $G = e^{-t} \tilde{G}$ which gives
\be
\frac{d}{dt} \det G = e^{-Nt} \frac{d}{dt} \det \tilde{G} - N \det G.
\label{eq:tilde}
\ee
Now the time-derivative in the first term on right-hand side can be written as 
\ber
\nonumber
\frac{d}{dt} \left | \begin{array}{cccc}
\tilde{G}_{11} & \tilde{G}_{12} & ...& \tilde{G}_{1N} \\
\tilde{G}_{21} & \tilde{G}_{22} & ...& \tilde{G}_{2N} \\
...            &   ...          & ...& ...\\
\tilde{G}_{N1} & \tilde{G}_{N2} & ... &  \tilde{G}_{NN}  
\end{array}
\right |
 = \left | \begin{array}{cccc}
{\dot {\tilde{G}}}_{11} & {\dot {\tilde{G}}}_{12} & ...& {\dot {\tilde{G}}}_{1N}
 \\
\tilde{G}_{21} & \tilde{G}_{22} & ...& \tilde{G}_{2N} \\
...            &   ...          & ...& ...\\
\tilde{G}_{N1} & \tilde{G}_{N2} & ... &  \tilde{G}_{NN}  
\end{array}
\right |
\\
+ \left | \begin{array}{cccc}
\tilde{G}_{11} & \tilde{G}_{12} & ...& \tilde{G}_{1N} \\
{\dot {\tilde{G}}}_{21} & {\dot {\tilde{G}}}_{22} & ...&{\dot {\tilde{G}}}_{2N} \\
...            &   ...          & ...& ...\\
\tilde{G}_{N1} & \tilde{G}_{N2} & ... &  \tilde{G}_{NN}  
\end{array}
\right |
 + ...+ \left | \begin{array}{cccc}
\tilde{G}_{11} & \tilde{G}_{12} & ...& \tilde{G}_{1N} \\
\tilde{G}_{21} & \tilde{G}_{22} & ...& \tilde{G}_{2N} \\
...            &   ...          & ...& ...\\
{\dot {\tilde{G}}}_{N1} & {\dot {\tilde{G}}}_{N2} & ... & {\dot
{\tilde{G}}}_{NN} 
\end{array}
\right |
\label{eq:gdot}
\eer
The $i$-th row of $\tilde{G}$ is $\left \{ {\tilde{G_{i1}}}, {\tilde{G_{i2}}},
...{\tilde{G_{iN}}} \right \} = \left \{ {\tilde{F}_{n(i,1)}}(x_i -y_1),
{\tilde{F}_{n(i,2)}}(x_i-y_2),...,{\tilde{F}_{n(i,N)}}(x_i-y_N) \right \} $ 
where we have used the notation $ F = e^{-t} \tilde {F}$.

To find the time-derivative of these elements of the $i$-th row 
 we use the identity ${\dot{F}}_n (x)=-F_n(x)+F_{n-1}(x-1)$ from which it follows that
${\dot{\tilde{F}}}_n(x) ={\tilde F}_{n}(x-1)$. Therefore, taking the time derivative generates
the row
 $ \left \{ {\tilde{F}_{n(i,1)}}(x_i-1 -y_1),
{\tilde{F}_{n(i,2)}}(x_i-1-y_2),...,{\tilde{F}_{n(i,N)}}(x_i-1-y_N) \right \}
$. Thus each term on the r.h.s. of (\ref{eq:gdot}) contributes
 to one term in the
summation in Eq. (\ref{eq:qqmaster}). From Eq. (\ref{eq:tilde}) it follows
therefore that $\det G$ satisfies Eq. (\ref{eq:qqmaster}). 

Now consider the case
when one nearest neighbor pair is present: $x_j=x_{j-1} +1$. If the $(j-1)$-th
particles is $A$ and the $j$-th one is $B$, then the master equation becomes
\ber
\nonumber
\frac{d}{dt} P^{Q|Q} \left (\bfx ;t|\bfy ;0 \right ) = 
\sum_{i \neq j} P^{Q|Q} \left (x_1,x_2,...,x_{i-1},
x_{i}-1,x_{i+1},...,x_N;t|\bfy ;0 \right ) 
\\
- N P^{Q|Q} \left
(x_1,x_2,...,x_N;t|y_1,\bfy ;0 \right ). 
\eer
which gives rise to the boundary condition 
\be
P^{Q|Q} \left ( x_1,x_2,...,x_{j-1},x_{j-1},x_{j+1},...,x_N ;t
|\bfy ;0 \right ) =0
\ee
to be satisfied at all times.
Since there is an $AB$ pair  at $(j-1), j$, from the definition
of $n(i,j)$ it follows that $n(i,j-1)=n(i,j)$ $\forall i$. Hence in the
determinantal representation of $ P^{Q|Q} \left ( x_1,x_2,...,x_{j-1},x_{j-1},x_{j+1},...,x_N ;t
|\bfy ;0 \right )$ the $(j-1)$-th and $j$-th row becomes identical
and the determinant vanishes. If the $(j-1)$-th particle is $B$ and the $j$-th
one is $A$ then the boundary condition is 
\ber
P^{Q|Q} \left ( x_1,x_2,...,x_{j-1},x_{j-1},x_{j+1},...,x_N ;t
|\bfy ;0 \right ) = \\ \nonumber 
P^{Q|Q} \left ( x_1,x_2,...,x_{j-1},x_{j-1}+1,x_{j+1},...,x_N ;t
|\bfy;0 \right )
\eer
which becomes
\ber
\nonumber
  \left | \begin{array}{ccc}
... & ... & ... \\
F_{n(j-1,1)}(x_{j-1} -y_1 ) & F_{n(j-1,2)}(x_{j-1} -y_2) & ... \\
F_{n(j-1,1)+1}(x_{j-1} -y_1 ) & F_{n(j-1,2)+1}(x_{j-1} -y_2) & ...\\
... & ... & ...
\end{array}
\right | = 
\\
\left | \begin{array}{ccc}
... & ... & ... \\
F_{n(j-1,1)}(x_{j-1} -y_1 ) & F_{n(j-1,2)}(x_{j-1} -y_2) & ... \\
F_{n(j-1,1)+1}(x_{j-1}+1 -y_1 ) & F_{n(j-1,2)+1}(x_{j-1}+1 -y_2) & ...\\
... & ... & ...
\end{array}
\right |
\eer
which can be proved using $ F_{n(j-1,i)+1}(x_{j-1} +1 -y_i) =
F_{n(j-1,i)+1}(x_{j-1}-y_i) - F_{n(j-1,i)}(x_{j-1}-y_i) $. 
If $(j-1)$-th and $j$-th particles are $AA$ or $BB$, then the corresponding
boundary condition becomes
\ber
P^{Q|Q} \left ( x_1,x_2,...,x_{j-1},x_{j-1},x_{j+1},...,x_N ;t
|\bfy ;0 \right ) = \\ \nonumber 
P^{Q|Q} \left ( x_1,x_2,...,x_{j-1},x_{j-1}+1,x_{j+1},...,x_N ;t
|\bfy ;0 \right )
\eer
which is easily shown to be satisfied using the identity $F_n(x+1)=F_n(x)
-F_{n-1}(x)$. 

Finally we have to show that $\det G$ satisfies the correct initial condition. To this
end note first the 
identity $F_n(x;t=0)=\delta_{x,0} $ which follows from the definition of 
$F_n(x;t)$ in (\ref{eq:fnx}). This means that for $x_1 > y_1$ all the
elements of the first column must vanish: $G_{i1}=0$ for $t=0$ which yields
$det G=0$. For $x_1=y_1$ one
has $G_{11}=1$ and  $G_{i1}=0$ $\forall i \neq 1$, since $x_i >y_1$ for $i>1$.
Then $\det G$ becomes equal to $ \delta_{x_1,y_1} \det G^{(1)}$ 
where the matrix $G^{(1)}$ is
obtained from $G$ after omitting the first row and the first column. Next, we
assume $x_2 > y_2$ and using the  above identity, we have $G^{(1)}_{i1} =0$
$\forall i$ and for $x_2=y_2$ $G^{(1)}_{i1}=\delta_{i,1} $. It follows
therefore that $\det G=\delta_{x_1,y_1} \delta_{x_2,y_2} \det G^{(2)}$, with
$G^{(2)}$ being an $(N-2)$ dimensional matrix which results from omitting
first two rows and columns of $G$. Repeating the above procedure $N$ times,
one finds 
\be
\det G   \vert_{t=0}= \prod_{i=1}^N \delta_{x_i,y_i}.
\ee   
This completes the proof of the theorem stated above. 


\section{Geometrical treatment of the Bethe Ansatz solution}
\label{sec:dia}

In this section, we use the geometrical interpretation of the Bethe Ansatz
introduced in~\cite{dia1,dia2}  to analyze the entangled systems of allowed
and forbidden trajectories of different species of particles. This approach
serves as a heuristic alternative to the explicit rigorous solution of the master equation. 

To
illustrate this approach let us first consider the single-species TASEP with
$N=2$. The solution of the master equation in this case is~\cite{Schu97} 
\be
F_0 (x_1 - y_1)F_0 (x_2-y_2) - F_1(x_2-y_1)F_{-1}(x_1-y_2).
\label{eq:2tasep}
\ee
 Instead of  the TASEP  let us now consider two  vicious random walkers (VRW) where
two particles perform biased random walks on a lattice and if they jump on
each other, they annihilate. It is easy to verify
that the solution of master equation in this case is
\be
 F_0 (x_1 - y_1)F_0(x_2-y_2) - F_0(x_2-y_1)F_{0}(x_1-y_2).
\label{eq:2vrw}
\ee
 Notice that in both the TASEP and the VRW, the
first term is the product of two Poisson processes, which represents two
non-interacting random walkers starting from $y_1, y_2$ and reaching
$x_1,x_2$, respectively. This term includes all possible trajectories for the
pair of particles. For the VRW, however,  trajectories are not allowed to
intersect as they will result in annihilation. So one must subtract those
forbidden trajectories where there exists at least one intersection. This is
obtained by considering two  non-interacting random walkers again, one
starting from $y_1$, reaching $x_2$ and the other starting from $y_2$ and
 reaching $x_1$. Since $x_1 < x_2$ and $y_1 < y_2$ their trajectories must
cross at least once and therefore (\ref{eq:2vrw}) yields the correct transition
probability. This observation follows also directly from the 
free-fermion representation
of this process, see \cite{lush,Schu01,mb1,mb2,Schu95} 
for a detailed treatment.

\begin{figure}
\includegraphics[scale=0.5,angle=0]{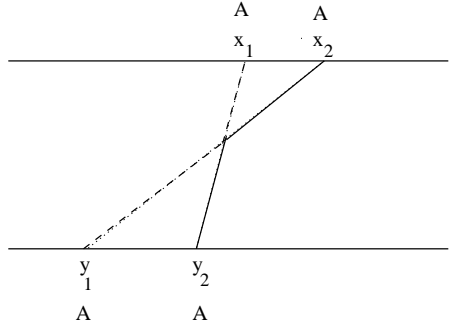}
\caption{Mean trajectories in the single-species TASEP. When trajectories of two particles collide
they reflect off each other. Here the horizontal line at the top shows the 
configuration of the system at time $t$ and that at the bottom shows the 
initial configuration.}
\label{fig:refl}
\end{figure}

For a TASEP, however, if the particles attempt to jump onto each other, they are
not annihilated but are reflected off each other (see Fig \ref{fig:refl}), i.e., the jump attempt is
unsuccessful. This means out of all the trajectories forbidden for the VRW
we allow those paths for the TASEP where particles collide and get reflected
off. As follows from the comparison of Eqs. (\ref{eq:2tasep}) and (\ref{eq:2vrw}), the
weight of this type of trajectories is  $F_0 (x_2-y_1)F_0(x_1-y_2) -
 F_1(x_2-y_1)F_{-1}(x_1-y_2)$. 

Now consider one $A$ particle and one $B$ particle. First consider the
probability that the initial sequence $\{AB\}$ remains unchanged at time $t$.
If the two particles jump on each other, they exchange
positions. Once the exchange has taken place, the particles are not allowed to
go back and the sequence remains $BA$ thereafter. This means that if we want
to compute the probability that even at time $t$ the sequence continues to be 
 $\{AB\}$, then we must not allow the particles to jump on each other. All
those trajectories where a jump attempt has been made, must be forbidden. In
other words, the two particles must behave like a pair of VRW's. This
explains the form of the solution in (\ref{eq:vr2}).

\begin{figure}
\includegraphics[scale=0.5,angle=0]{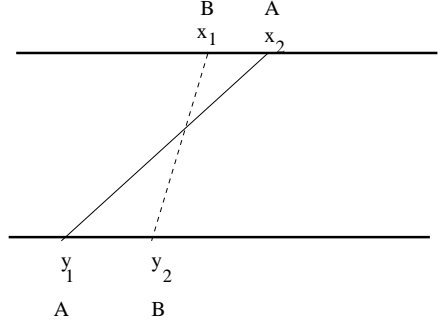}
\caption{Diagram showing interchange of an $A$ and $B$ particle}
\label{fig:inter}
\end{figure}

For the case when the $A$ and $B$ particles actually interchange and the
sequence becomes $BA$, one must consider only those paths where there exists 
a crossing, as shown in Fig \ref{fig:inter}.
 Note that this diagram can be obtained from Fig \ref{fig:refl}
 after `recoloring' the lines. Even for the single-species TASEP, where
particles are not allowed to interchange positions, one can reinterpret the
diagram \ref{fig:refl}
 as a process where two particles actually interchange positions and
then renamed after the interchange (corresponds to the exchange of the colors
of the lines). Hence Figures \ref{fig:refl} and \ref{fig:inter}
  are essentially identical and 
the weight of the allowed trajectories in this case is 
\be
F_0 (x_2-y_1)F_0(x_1-y_2) - F_1(x_2-y_1)F_{-1}(x_1-y_2)
\label{eq:v+}
\ee
 as derived above. Again, our rigorous solution in
(\ref{eq:vr2}) conforms to this argument.  

Thus we have developed an equivalent diagrammatic approach to interpret the
Bethe Ansatz solution presented in section \ref{sec:bethe}. This approach can
easily be generalized for higher values of $N$. There are three different
possibilities when two particles interact---when an $A$ interacts with a $B$ on
its right, they can behave like VRW or they can exchange. For any other
collision, the particles are reflected off each other (in Fig \ref{fig:refl} the
trajectory starting from $y_1$ is reflected backward while the other starting
from $y_2$ is reflected forward). For a set of $N$ particles, there are $N!$
number of possible diagrams obtained from connecting all $\{y_1,y_2,...,y_N\}$
with all $\{x_1,x_2,...,x_N\}$. Clearly, each diagram corresponds to one
particular permutation for the Bethe momentum variables $\{p_1,p_2,...,p_N\}$
with the conjugate position variables $\{x_1,x_2,...,x_N\}$. In a diagram, if
$y_j$ is joined to $x_i$, then the corresponding term in the solution must
have an $F$ function with argument $(x_i-y_j)$. To obtain the index
(subscript) of this $F$ function, count the number of intersections on the line 
joining $x_i$ and $y_j$. Each forward (backward) reflection corresponds to an
index -1(+1), each VRW collision corresponds to an index $0$ and each $AB$
interchange is represented as difference of two terms, as in Eq. (\ref{eq:v+}). 
For illustration, we provide an example below.

\begin{figure}
\includegraphics[scale=0.5,angle=0]{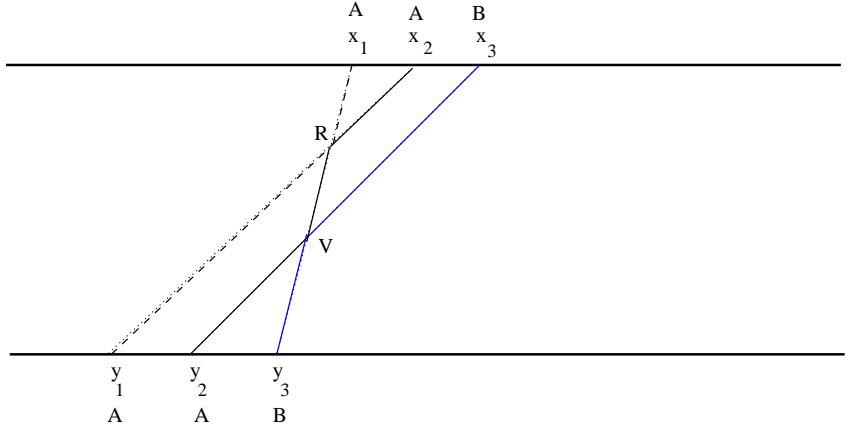}
\caption{The sequence $AAB$ remains same in time and the $AB$ pair must behave
like VRW.}
\label{fig:3vr}
\end{figure}

In Fig \ref{fig:3vr}
 the intersection marked $V$ is between an $AB$ pair and since they do
not interchange they must behave as VRW. So collision $V$ yields an index $0$
in the $F$ functions with arguments $(x_1-y_3)$ and $(x_3-y_2)$. The collision
$R$ is between two $A$ particles and hence of ordinary reflecting type, where
the line $(x_2-y_1)$ suffers a backward reflection and $(x_1-y_3)$ has a
forward reflection. Final weight of the diagram is
$F_0(x_3-y_2)F_{-1}(x_1-y_3) F_1(x_2-y_1)$. 

\begin{figure}
\includegraphics[scale=0.5,angle=0]{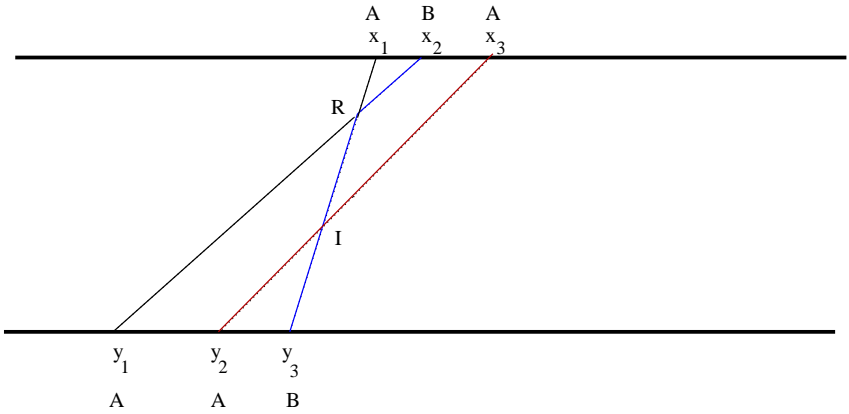}
\caption{Sequence $AAB$ changes to $ABA$ and in the collision marked $I$
the $AB$ pair actually interchange and collision $R$ is of reflecting type}
\label{fig:3inter}
\end{figure}

Consider another example. In Fig \ref{fig:3inter}
  in the collision $I$ the $A$ and $B$
interchange and in collision $R$ the line $(x_2-y_1)$ has a backward
reflection and $(x_1-y_3)$ has forward reflection. These taken together yields 
$F_1(x_2-y_1) \left [ F_{-1}(x_1-y_3)F_0(x_3-y_2) -
F_{-1}(x_3-y_2)F_0(x_1-y_3) \right ]$. Note that the part in the square
bracket has the same form as in Eq. (\ref{eq:v+}) but because $(x_1-y_3)$
suffers a forward reflection with $(x_2 - y_1)$, its index is lowered by
unity. 

Note that not all the $N!$ diagrams are always allowed. In the above example, 
when an initial sequence $AAB$ changed to $ABA$, then not all diagrams
are possible: Since we want the $B$ particle starting at $y_3$ to interact
with the other $A$ particles, we cannot have any diagram where $x_3$ is joined
to $y_3$, meaning no intersection lies on the $(x_3-y_3)$ line. In Bethe
Ansatz language this means that all permutations where $p_3x_3$ appears are absent
from the solution.

Thus these diagrams serve as an alternative way to interpret the terms in the
Bethe Ansatz solution. This approach also offers us an insight into the
theorem discussed in section \ref{sec:theorem}. We can now understand why the
determinant in \cite{Schu97} for the single species case gets modified into $\det G$ for the
two-species case obtained above. When only one type of particles is present, the
intersections in their trajectories result in only reflecting type collisions
and the corresponding index of the $F$ function is either $+1$ or $-1$. It can
be easily seen that the line $(x_i-y_j)$ picks up an index $(i-j)$ for the $F$
function, as a result of all these reflections. But when both $A$ and $B$
particles are present, and if we are interested in the case when  their
sequence do not change, then some of these collisions have to be of `vicious'
type which corresponds to an index $0$. For the line $(x_i-y_j)$ 
the number of such `vicious' collisions
is $n_{AB}$, the number of  $AB$ pairs between the $i$-th and
$j$-th particle. As a result, the corresponding index of the $F$ function is
not $(i-j)$ but $sgn(i-j)(|i-j|-n_{AB})$. This explains the structure of the
matrix $G$ in section \ref{sec:theorem}.

\section{Further Discussion}
\label{sec:last}

In this paper  we have used the Bethe Ansatz to solve the master equation of the
TASEP with first and second class particles for arbitrary initial conditions. 
We have also obtained a compact determinantal
representation of the solution for the case when the initial sequence of particles remains
unchanged. An equivalent geometrical approach developed in section
\ref{sec:dia} provides us with insight into the form of the determinant. 
We have seen that when the sequence does not change with time, then all the 
$AB$ pairs in the sequence must behave like VRWs. 

Interestingly the last 
conclusion remains valid even when more than two species of particles are
present in the system. For any initial sequence $Q$, consisting of any arbitrary
number of species (classes)
 of particles, if we are interested in the probability that the sequence does
not change until time $t$, then all the pairs, where a higher class particle
has a lower class particle to its right, must behave like VRWs. All the
trajectories where they jump onto each other, must therefore be forbidden
because once they interchange positions, they cannot go back again.
Generalizing the geometrical approach of section \ref{sec:dia}, it is easy to
see that even in this case the diagrams contain two different types of
intersection points: one of vicious type (corresponding to an index $0$)
 and the other of reflecting type (corresponding to an index $\pm 1$). For
any given line $(x_i-y_j)$, there will be an $F$ function as before, with the
argument $(x_i-y_j)$ and index $n(i,j)= sgn(i-j)(|i-j|-n_v)$ where $n_v$
denotes the number of `vicious pairs' (where a higher class particle has a
lower class particle on its right) between $i$-th and $j$-th member of the
sequence. Thus using this approach one can write down a determinantal
representation for the solution of the master equation for a general multi-class TASEP
where the sequence of various species of particles remains unchanged.

In the special case of a sequence where the first particle from the
left is of class $1$, the second particle is of class $2$, the third particle
is of class $3$, ...,the $N$-th particle is of class $N$,  it follows from the above
argument that all the pairs in this sequence behave like VRW's if we
consider the probability that the sequence is preserved until time $t$. This
probability is nothing but $\det F^0$ where $F^0$ is a matrix with $(ij)$-th
element ${F^0}_{ij} = F_0 (x_i-y_j)$. If on the other hand the initial 
sequence is
reversed when the first class particle  becomes the rightmost particle and the
$N$-th class particle becomes the leftmost particle, then there are no vicious
pairs left in the system and the solution for the single species TASEP
remains valid. Thus depending on the labels of the particles in the initial 
sequence, one can have TASEP solution or VRW solution. 

In this context it is interesting to discuss another recent 
result on the crossover from VRW to TASEP. In~\cite{semi} a model for 
semi-vicious walkers was introduced that interpolates between VRW and 
TASEP, having the two as limiting cases. In this model two particles
annihilate if they jump on each other, but this jump attempt to an occupied
site takes place with a reduced probability. Note that for VRW's the probability
to jump on an occupied site is same as that for an empty site, while for the TASEP
jumping on an occupied site is prohibited. Thus by varying the probability 
to jump on an occupied site one can go from one limiting case to the other
and the model for semi-vicious walkers serves an intermediate between these
two limits. It was shown that the survival probability
for $N$ particles can be described by a scaling form that characterizes the
transition from VRW to TASEP~\cite{semi}. In contrast we consider here a 
multi-species exclusion process and have argued in the previous
paragraph that if each of the $N$ particles belongs to a distinct class, then 
it is possible to have a VRW or a TASEP solution depending on the label of
the particles in the initial sequence. 
 For a general sequence, under the condition that until time
$t$ the sequence remains unchanged, some of the particles behave like TASEP
and some behave like VRW. Thus both these aspects are simultaneously present
in our model. In the case when the number of vicious interfaces $n_v=N-1$,
 all the particles behave like VRW and the
solution of the master equation becomes $\det F^0$, while for $n_v=0$ all the
particles behave as TASEP particles. For intermediate
values of $n_v$ both these limiting behaviors are present and some pairs
behave like TASEP and some like VRW and the determinantal solution of the
master equation is modified accordingly.

For the general two-species problem, i.e. when the initial sequence of species changes because of interchange of $A$
and $B$ particles, it is an interesting question whether the Bethe ansatz solution
presented in this work allows for a compact representation in terms of determinants.
At this point we do not have a complete answer to
this question. We have been able to obtain compact representations for
certain special cases. For example, if the final sequence $Q'$ results from a
single $AB$ interchange from the initial sequence $Q$, such that the number of
$AB$ pairs in $Q'$ is less than that in $Q$, then we
have been able to prove that the probability $P^{Q'|Q} = P^{Q'|Q'}-P^{Q|Q}$
can be written as a difference of two determinants. The existence of similar compact
representation for any number of interchanges and for the general
multi-class TASEP is an open problem for future investigation.

\section{Acknowledgments}
We acknowledge useful discussions with V.B. Priezzhev, T. Sasamoto, 
P.L. Ferrari, A. R\'akos, P. Gon\c{c}alves. 
Financial support from Deutsche Forschunsgemeinschaft is gratefully acknowledged.
  
\appendix
\section{Verification of Yang-Baxter Criterion} 
\label{app:yb}

In this appendix, we discuss the validity of Yang-Baxter equation for 
$N=3$ case. Consider the r.h.s. of  Eq. (\ref{eq:bethe3}).
The matrices $\Sigma_{12}^{3}(p_j,p_k)$ and $\Sigma_{23}^{3}(p_j,p_k) $
represents elementary permutations of the momenta variables. For example,
$\Sigma_{12}^{3}(p_j,p_k)$ represents the interchange of the momentum
 variables $p_j$ and $p_k$ which are coupled to $x_1$ and $x_2$,
 respectively. Each term in Eq. (\ref{eq:bethe3}) corresponds to one particular
permutation ${\cal P}$ of the momenta and the product of the $\Sigma$ matrices
 represents
decomposition of ${\cal P}$ into a series of elementary permutations. This
decomposition need not be unique for all  ${\cal P}$. Consider, for example, 
the last term in Eq. (\ref{eq:bethe3}), where the momenta variables coupled 
to $x_1,x_2,x_3$, are, respectively, $p_3,p_2,p_1$. 
 Starting from a permutation  $p_1,p_2,p_3$ (as in the first
term on the r.h.s. of Eq. (\ref{eq:bethe3}), to reach the arrangement
$p_3,p_2,p_1$, one can first interchange $(p_2,p_3)$, then $(p_1,p_3)$ and
again $(p_1,p_2)$. Accordingly, the $\Sigma$-product is  
$\Sigma_{23}^{3}(p_1,p_2)\Sigma_{12}^{3}(p_1,p_3) \Sigma_{23}^{3}(p_2,p_3)$. 
But the same permutation can also be reached by first interchanging
$(p_1,p_2)$, then $(p_1,p_3)$ and finally $(p_2,p_3)$. In that case, the
corresponding matrix will be 
$\Sigma_{12}^{3}(p_2,p_3)\Sigma_{23}^{3}(p_1,p_3) \Sigma_{12}^{3}(p_1,p_2)$. 
In order for the system to be integrable, one must satisfy the Yang-Baxter criterion
\be
\Sigma_{12}^{3}(p_2,p_3)\Sigma_{23}^{3}(p_1,p_3) \Sigma_{12}^{3}(p_1,p_2) = 
\Sigma_{23}^{3}(p_1,p_2)\Sigma_{12}^{3}(p_1,p_3) \Sigma_{23}^{3}(p_2,p_3).
\label{eq:yb}
\ee
(We remark that the slightly different Yang-Baxter equation for the scattering matrix $S$
is obtained from this croterion through the relation $S=\Sigma P$ where $P$ is the 
permutation operator). Now, using the definitions
 $\Sigma_{12}^{3}(p_j,p_k) \equiv \Sigma (p_j,p_k) \otimes {\mathbb 
I}_2 $ and $\Sigma_{23}^{3}(p_j,p_k) \equiv {\mathbb I}_2 \otimes
\Sigma (p_j,p_k) $, we have 
\be
\Sigma_{12}^{3}(p_i,p_j) =\left ( 
\begin{array}{cccccccc}
S_{jk} & 0& 0& 0& 0& 0& 0& 0\\
0 & S_{jk}& 0& 0& 0& 0& 0& 0 \\
0 & 0 & -1& 0& 0& 0& 0& 0 \\
0 & 0& 0& -1& 0& 0& 0& 0 \\
0 & 0 & T_{jk} & 0& S_{jk} & 0& 0& 0 \\
0 & 0 & 0 & T_{jk} & 0& S_{jk} & 0& 0 \\
0 & 0& 0 &  0& 0 & 0& S_{jk} & 0 \\
0 & 0& 0 &  0& 0 & 0& 0 & S_{jk} 
\end{array}
\right ) 
\ee 
and 
\be
\Sigma_{23}^{3}(p_i,p_j) =\left ( 
\begin{array}{cccccccc}
S_{jk} & 0& 0& 0& 0& 0& 0& 0\\
0 & -1& 0& 0& 0& 0& 0& 0\\
0 & T_{jk} & S_{jk} & 0& 0& 0& 0& 0\\
0 & 0& 0& S_{jk}& 0& 0& 0& 0\\
0 & 0& 0& 0 & S_{jk}& 0& 0& 0\\
0 & 0& 0& 0 & 0& -1& 0 & 0 \\
0 & 0& 0& 0 & 0& T_{jk}& S_{jk}& 0 \\
0 & 0& 0& 0 & 0&0 & 0& S_{jk} 
\end{array}
\right )
\ee
where $S_{jk} = - (1-e^{ip_j})/(1-e^{ip_k})$ and
$T_{jk}=-(e^{ip_k}-e^{ip_j})/(1-e^{ip_k})$. Using these forms it is
straightforward to verify Eq. (\ref{eq:yb}).  
 
We remark that the nested Bethe ansatz can be also be used for treating 
second-class particles in the partially asymmetric simple
exclusion process (PASEP) where particles jump to the right or left with  non-zero rates \cite{arita}. 
However, even for the single-species PASEP there is no known determinantal representation 
of the transition probabilities. 


\end{document}